\documentclass[letterpaper,twocolumn,english,prl,aps,superbib,tightenlines,floatfix]{revtex4-1}
\usepackage[T1]{fontenc}
\usepackage[latin9]{inputenc}
\usepackage{amsmath}
\usepackage{graphicx}
\usepackage{amssymb}

\usepackage{babel}

\begin{document}
\title{Ionic Memcapacitive Effects in Nanopores}
\author{Matt Krems$^{1}$, Yuriy V. Pershin$^{2}$, Massimiliano Di Ventra$^{1}$}
\affiliation{ $^{1}$Department of Physics, University of California, San Diego, La Jolla, CA 92093
\\ $^2$Department of Physics and Astronomy and USC Nanocenter, University of South Carolina, Columbia, SC 29208}
\date{\today}

\begin{abstract}
Using molecular dynamics simulations, we show that, when subject to a periodic external electric field, a nanopore in ionic solution acts as a capacitor with memory (memcapacitor) at various frequencies and strengths of the electric field. Most importantly, the hysteresis loop of this memcapacitor shows both negative and diverging capacitance as a function of the voltage. The origin of this effect stems from the slow polarizability of the ionic solution due to the finite mobility of ions in water. We develop a microscopic quantitative model which captures the main features we observe in the simulations and suggest experimental tests of our predictions. We also suggest a possible memory mechanism due to the transport of ions through the nanopore itself, which may be observed at small frequencies. These effects may be important in both DNA sequencing proposals using nanopores and possibly in the dynamics of action potentials in neurons.
\end{abstract}
\maketitle

A nanopore is an aperture of nanoscale dimensions across an insulating membrane. This way, if immersed in an ionic solution, ions can
enter the nanopore, and if subject to an electric field of given strength, they may enter it
from one side of the opening and emerge on the other side. The
membrane can be made of either biological or solid-state materials. Nanopores are
ubiquitous in biological systems as they regulate the flow of ions across cell membranes
(of, e.g., neurons). At present, they are also actively investigated for potential applications
in DNA sequencing~\cite{zwolak08,branton08}. Despite its apparent simplicity, ion dynamics in nanopores is
far from trivial. An example of this is provided by the recently predicted phenomenon of
``quantized ionic conductance'', namely the presence of current steps occurring at effective pore radii
that match the radii of the water hydration layers that form around the ions~\cite{zwolak09}.

Here, we predict another intriguing property of ion dynamics in
nanopores, when the latter are subject to a periodic external
electric field. In particular, we show that the electric field
forces ions to accumulate at the two surfaces of the nanopore membrane,
creating an effective capacitor. However, this capacitor shows
interesting features as a function of the frequency of the field,
namely a hysteresis loop of the capacitance (memory-capacitance or {\it memcapacitance} for
short~\cite{diventra09}) as a function of
voltage which diverges at zero voltage and displays
negative-capacitance properties. Memcapacitors have been theoretically formulated in Ref.~\cite{diventra09} and belong
to the wider class of memory-circuit elements, which includes also memristors and meminductors, namely resistors and
inductors with memory. Following Ref.~\cite{diventra09} we define a voltage-controlled
memcapacitive system by the set of equations
\begin{eqnarray}
Q(t)&=&C\left(x,V,t \right)V(t) \label{Ceq1} \\
\dot{x}&=&f\left(x,V,t\right) \label{Ceq2}
\end{eqnarray}
where $Q(t)$ is the charge on the capacitor at time $t$, $V(t)$
is the voltage across it, and $C$ is the time-dependent {\it memcapacitance} which
depends on the internal state of the system described by a set of $n$ state variables $x$, with $f$ a continuous
$n$-dimensional vector function. Using
all-atom molecular dynamics (MD) simulations, we indeed demonstrate that
memory characteristics of nanopores in solution originate from the finite mobility of ions in water
with consequent slow polarizability of the ionic solution. We also
develop a simple microscopic model that captures the main properties
observed in the simulations and allows us to extend our results to
regimes beyond the reach of MD simulations. Additionally, we use an equivalent circuit formulation to discuss how ionic transport through the nanopore itself may lead to an additional memory mechanism.
We also propose ways our
predictions could be tested with available experimental
capabilities.

\begin{figure} [b]
\centering
\includegraphics[width=7.5cm]{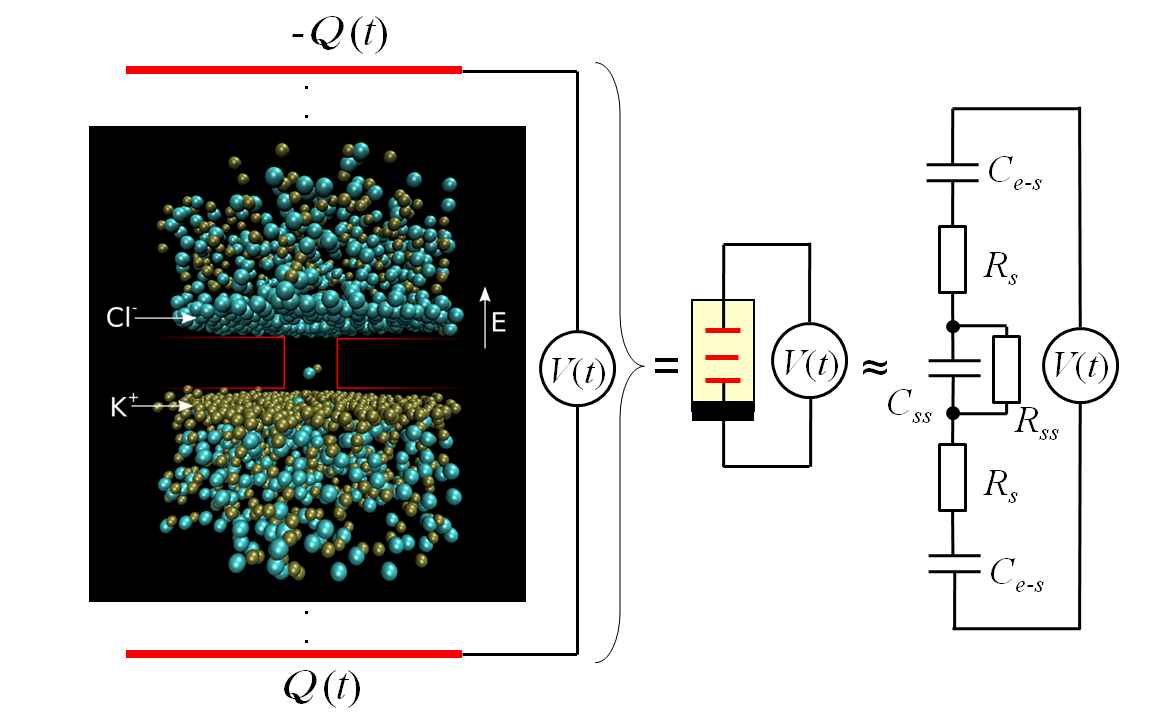}
\caption{(Color online) Left: a snapshot
of the molecular dynamics geometry at a time when a buildup of
charges of the opposite sign on each side of the nanopore is
observed due to a finite electric field $E$. The nanopore membrane is located at
the center (represented by the red lines) and water is not visible.  Top and bottom horizontal
red lines represent electrodes (holding plate charges $\pm Q(t)$) of
the suggested experimental set-up needed to observe the predicted
memcapacitive effects. Right:
simplified equivalent circuit model. \label{fig:schematic}}
\end{figure}

{\it Ionic memcapacitors} --- In order to show that a nanopore in
solution indeed acts as a memcapacitive system (Eqs.~(\ref{Ceq1})
and~(\ref{Ceq2})), let us be more specific and consider the
experimental situation as depicted in Fig.~\ref{fig:schematic}
which we suggest as a way to observe the effects predicted here.
An external (time-dependent) voltage $V(t)$ is applied to
electrodes which form an effective capacitor with the ionic
solution and the nanopore at its interior. In this paper we consider a typical setup for, e.g., DNA
sequencing devices~\cite{lagerqvist06,lagerqvist07, krems09} with
KCl ionic solution and a Si$_3$N$_4$ nanopore 25 \AA~thick and 88
\AA~wide with a 7 \AA~radius cylindrical hole drilled through the
center of the membrane. However, the specific dimensions of the
pore are irrelevant for the overall conclusions of this paper and similar
considerations would apply also to biological pores. When a global
time-dependent bias is applied to the system, the ionic solution
is polarized and ions are forced to accumulate at the surfaces of
the pore (see Fig.~\ref{fig:schematic}). First, we
consider relatively short time scales when ion transport across
the nanopore is not significant. (We will discuss towards the end of the paper the effect of ion transport across the pore at longer times.)
This implies that the same (and
opposite in sign) charge, $Q(t)$, that accumulates on the surfaces
of the pore also accumulates on the plates of the capacitor.

In the simplest approximation, the electrical properties of the total
nanopore system can be approximated by an equivalent circuit model as shown
in Fig.~\ref{fig:schematic}, where capacitors and resistors
(generally, non-linear) represent different parts of the system. Here, $C_{e-s}$ denotes the
external electrode-solution capacitance, $R_s$ is the resistance of the solution,
$R_{ss}$ ($\gg R_s$) is the resistance of the ion current through the pore, and $C_{ss}$ is the capacitance of the membrane.
The total voltage drop is then given by
\begin{equation}
2\frac{Q}{C_{e-s}}+2R_{s}\frac{{\textnormal d}Q}{{\textnormal d}t}+\frac{Q}{C_{ss}}=V(t).
\label{voltage_drop}
\end{equation}
Since we are interested in the properties of the pore and solution only,
we envision the experimental set-up so that $C_{e-s}\gg C_{ss}$. The equation for the charge then takes the form
\begin{equation}
\frac{{\textnormal d}Q}{{\textnormal d}t}=\frac{V(t)}{2R_{s}}-\frac{Q}{2R_sC_{ss}}.
\label{charge_dynamics}
\end{equation}
Eq. (\ref{charge_dynamics}) describes a relaxation of $Q$ towards
$V(t)/(2R_s)$ with a relaxation time $2R_sC_{ss}$. Generally,
due to the finite mobility of ions in
water, the charge on the external plates - and hence on the pore -
is slow to respond to changes of the bias upon varying $V(t)$.
This means that $V(t)$ may be zero when $Q(t)$ is finite and vice
versa. Therefore, the capacitance of this whole system,
$C=Q(t)/V(t)$, depends on the applied voltage history, shows
divergences and acquires negative values at specific times, similar to what has been recently predicted in
some solid-state memcapacitors~\cite{martinez09a}. Since
these memory effects can be viewed as originating from the
history-dependent permittivity of the ionic solution, following
the definition of Ref.~\cite{diventra09} (Eqs.~(\ref{Ceq1})
and~(\ref{Ceq2})), a nanopore in ionic solution is indeed a
memcapacitive system.

{\it Results and discussion} --- Let us now demonstrate these memcapacitive
effects from a microscopic point of view. For this we employ all-atom MD
simulations using NAMD2~\cite{phillips05} and investigate the
response of the system to external ac- and dc-electric fields. According to our previous
discussion, in order to calculate the total capacitance of the system we only need the net charge
that accumulates at the nanopore surface. We then employ periodic boundary conditions in both the
direction perpendicular and parallel to the external field. On each side of the pore, we place a
50 \AA~long section of water with a 1 M solution of a homogenous,
random distribution of potassium and chlorine ions. The CHARMM27
force field~\cite{foloppe99,mackerell99} is used for the
interaction of water and ions while quantum mechanical parameters~\cite{wendell92} are used for the
Si$_{3}$N$_{4}$ pore~\footnote{The Si$_{3}$N$_{4}$ atoms are harmonically
confined in order to mimic the dielectric properties of
Si$_3$N$_4$. A 1 fs time step is used and the system temperature
is kept at room temperature with a Langevin dampening parameter of
0.2 ps$^{-1}$ in the equations of motion~\cite{aksimentiev04}. The
van der Waals interactions are gradually cut off starting at 10
\AA$\,$ from the atom until reaching zero interaction 12 \AA$\,$
away. The energy was initially minimized in 1000 time steps and
then equilibrated for 1 ns with a zero electric field.}.

\begin{figure}
\centering
\includegraphics[width=7.5cm]{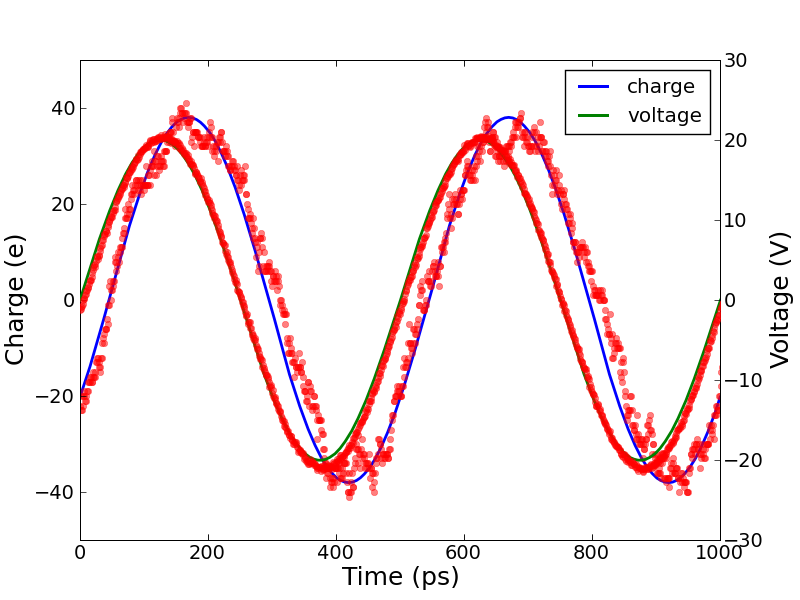}
\caption{(Color online) Net charge on the surface of the nanopore (equal to system's plate charge) and voltage across the nanopore (which is very close to the external voltage, solid line) plotted as a
function of time. The solid line for the charge corresponds to the model we discuss in the text. One can clearly see that the net charge lags behind in time
with respect to {\em both} the external voltage and the voltage at the pore.
\label{fig:fit}}
\end{figure}

We have performed extensive MD simulations of the process of
polarization of the ionic solution in proximity to the nanopore
membrane. As a general
observation, within the parameters used, a delay between the
formation of ionic polarization and {\em both} the external ac-electric field and
the voltage across the pore has been
observed. This is evident in Fig.~\ref{fig:fit} where the net charge
of ions on one side of the nanopore is plotted as function
of time together with the voltage drop across the membrane. The
latter is obtained by doing a full electrostatics calculation from
MD simulations~\cite{aksimentiev05}. At this point we remark that
in all our simulations we have found that the net charge induced
by the external field is always located within 10 \AA~from the
membrane surface and that the voltage drop across the membrane is
very close in value to the external applied voltage, whatever the
strength and frequency of the latter. This last observation is
consistent with previous studies~\cite{zwolak08}.

\begin{figure}
\centering
\includegraphics[width=7.5cm]{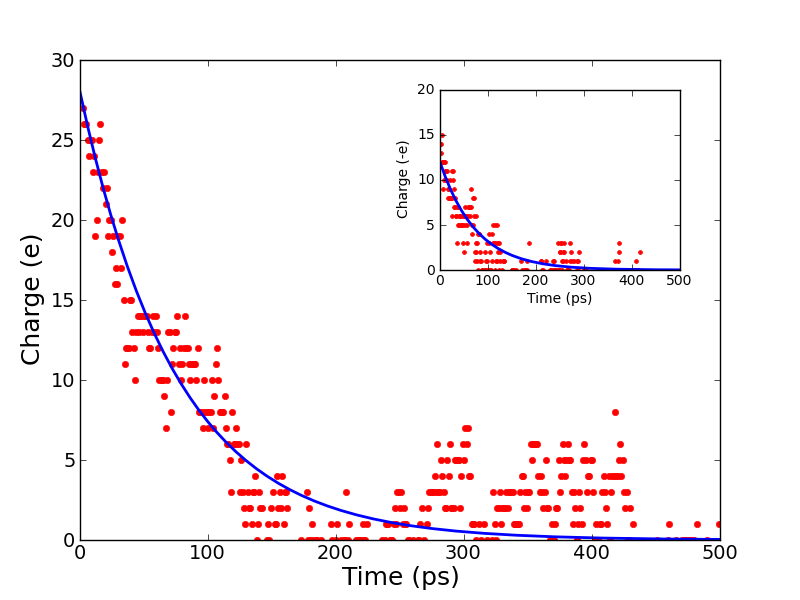}
\caption{(Color online) Net charge of the positively charged side of the
the capacitor vs. time when a constant 20 V applied
voltage responsible for the accumulation of the charges on the surface of the pore is turned off. A decay timescale of 75 ps is obtained by fitting the
simulation results with an exponential decay curve, i.e.,
$Q(t)=Q_{max}e^{-t/\tau}$. Due to the
nearly equal mobilities of K$^+$ and Cl$^-$~\cite{koneshan98}, we find
similar results for the negatively charged side and different field strengths as well. This is shown in the inset with initial voltage of 5 V.
\label{fig:decay}}
\end{figure}

Next, we fit these MD simulation results employing the equivalent circuit model given by
Eq. (\ref{charge_dynamics}). Using microscopic parameters, Eq. (\ref{charge_dynamics}) can be rewritten as
\begin{equation}
 \frac{{\textnormal d}Q(t)}{{\textnormal d}t} = 2e A\mu n E(t) - \frac{Q(t)}{\tau}, \label{eqQ}
\end{equation}
where $A$ is the area of the membrane surface, $\mu$ is the ion
mobility which is similar for both types of ions and we take to
be $7.12\times 10^{-8}\:m^2V^{-1}s^{-1}$~\cite{koneshan98}, $e$ the ion charge, $n$
the density of ions in the bulk, and $\tau=2R_sC_{ss}$ is the
relaxation time. The factor of 2 takes into account the conductivity of both channels (K$^+$ and Cl$^-$). Using dc-field simulations, we were able to
extract $\tau$. We do this by first applying a constant electric
field (with the same magnitude as the amplitude of our ac-field
simulations) which forces ions to accumulate in the vicinity of
the membrane as it is shown in Fig. \ref{fig:schematic}. The total
net charge within 10 \AA~from the nanopore surface is integrated
to obtain the capacitor charge~\footnote{Here, we do not consider
possible protonation of water which may cause the number of mobile
charges to increase and thus change the number of charges that
buildup on the pore, effectively increasing the ionic
concentration.}. Then, we turn off the field and monitor the decay
of this charge back into the bulk (see Fig.~\ref{fig:decay}). From
this we obtain the relaxation time $\tau$. We find that this time
is slightly dependent on the amplitude of the initially applied
electric field, becoming slightly longer as the applied field
decreases. For voltages as low as 5 V across the membrane we get
from our simulations a relaxation time $\tau\sim
75$ps~\footnote{Smaller voltages give rise to a very noisy net
charge, and it is thus difficult to extract a relaxation time from
them.}.

\begin{figure}
\centering
\includegraphics[width=7.5cm]{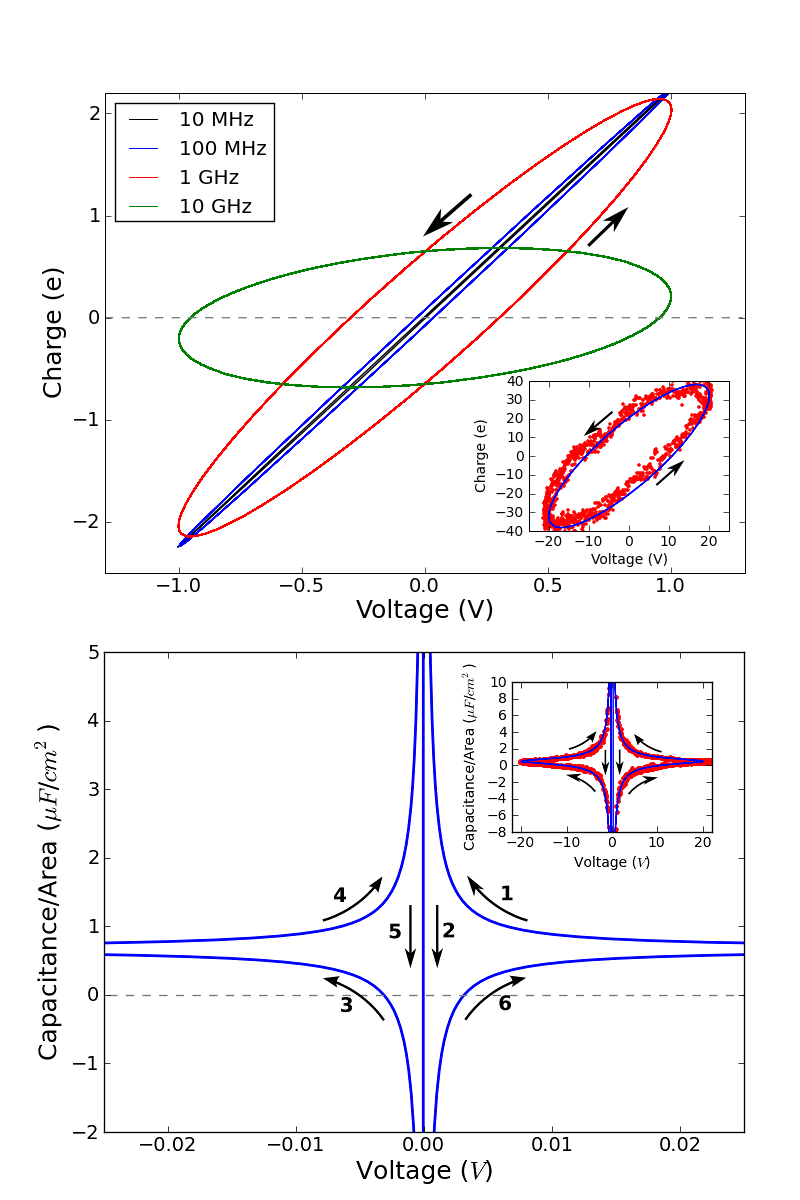}
\caption{(Color online) Upper panel: net charge $Q$ versus a
periodic voltage of amplitude $V_0=1$ V and different frequencies
as obtained from Eq.~\ref{longQ}. The inset shows the same
quantity for 2 GHz and $V_0=20$ V compared with the charge
obtained directly from our MD simulations. Lower panel:
capacitance versus a periodic voltage of frequency $f=$10 MHz and
amplitude $V_0=1$ V as obtained from Eq.~\ref{eq:C}. In the inset
we show the same quantity for $f=$2 GHz
 and $V_0=20$ V compared with the capacitance obtained directly from our MD simulations. It is seen that $C$ can be both negative and diverges as
the voltage approaches zero. The arrows indicate the direction the voltage is swept in time and the numbers show the order in which the trace is generated.
\label{fig:infiniteC}}
\end{figure}

Solving Eq.~(\ref{eqQ}) for $Q(t)$ with
an ac electric field $E(t)=E_0 \sin{(\omega t)}$ of amplitude $E_0$ and angular frequency $\omega=2\pi f$,
we obtain the long-time limit solution:
\begin{equation}
 Q(t) \stackrel{t \to \infty}{=} \frac{2A\mu en E_0 \sin{(\omega t)}}{\left(1+\frac{1}{\omega^2 \tau^2}\right)\omega^2 \tau} -
 \frac{2A\mu en E_0 \cos{(\omega t})}{\left(1+\frac{1}{\omega^2 \tau^2}\right)\omega}\label{longQ}
\end{equation}
which amounts to a sine function with a phase shift, namely $Q(t) = Q_{0} \sin{(\omega t - \delta)}$,
where
\begin{equation}
 Q_{0} = \left[\left(\frac{2A \mu e n E_0}{\left(1+\frac{1}{\omega^2 \tau^2}\right)\omega}\right)^2 + \left(\frac{2A \mu e n E_0}{\left(1+\frac{1}{\omega^2 \tau^2}\right)\omega^2\tau}\right)^2\right]^{1/2}
\end{equation}
and $\delta=\tan^{-1}{(\omega\tau)}$.
Using the appropriate values of $A$, $n$, and $\tau$ as obtained from the numerical simulations, this model agrees very well with the results of our MD simulations as it is
evident from Fig.~\ref{fig:fit}. In particular, it is clear that the net charge and voltage on the capacitor are phase shifted by the amount $\delta$. This means that when the voltage
is zero the charge on the capacitor is not necessarily zero, and vice versa. This is shown in Fig.~\ref{fig:infiniteC}(a) where the net charge is plotted as a function of the voltage across the capacitor. This gives rise to diverging and negative values of capacitance.

To see this, we write the external voltage across the whole system as $V(t) =
V_0 \sin{(\omega t)}$, with $V_0=-E_0 d$ where $d$ is the size of the simulation cell along the direction of the electric field. From $C=Q(t)/V(t)$ and~(\ref{longQ}) we then obtain the
long-time limit of the capacitance:
\begin{equation}
 C(t) \stackrel{t \to \infty}{=} \frac{2A\mu en}{\left(1+\frac{1}{\omega^2 \tau^2}\right)\omega^2 \tau d} - \frac{2A\mu en}{\left(1+\frac{1}{\omega^2 \tau^2}\right)\omega d} \cot{(\omega t)},
\label{eq:C}
\end{equation}
This represents the main result of our paper. It shows that the
capacitance of the whole nanopore system is history-dependent,
diverges when the voltage crosses its zero value, and acquires negative values
within a certain voltage range. This is plotted in
Fig.~\ref{fig:infiniteC} where both the charge on the capacitor
and the capacitance are plotted as a function of voltage across
the capacitor. From Fig.~\ref{fig:infiniteC} it is evident that at
low frequencies, this ionic memcapacitor behaves almost like a
regular capacitor. At higher frequencies, memory effects in the
capacitance manifest in a hysteresis loop characteristic of
memcapacitors. However, unlike typical memcapacitors that show a
pinched hysteretic loop~\cite{diventra09}, ionic memcapacitors
have a non-vanishing (diverging) zero-bias capacitance.

\begin{figure}
\centering
\includegraphics[width=7.5cm]{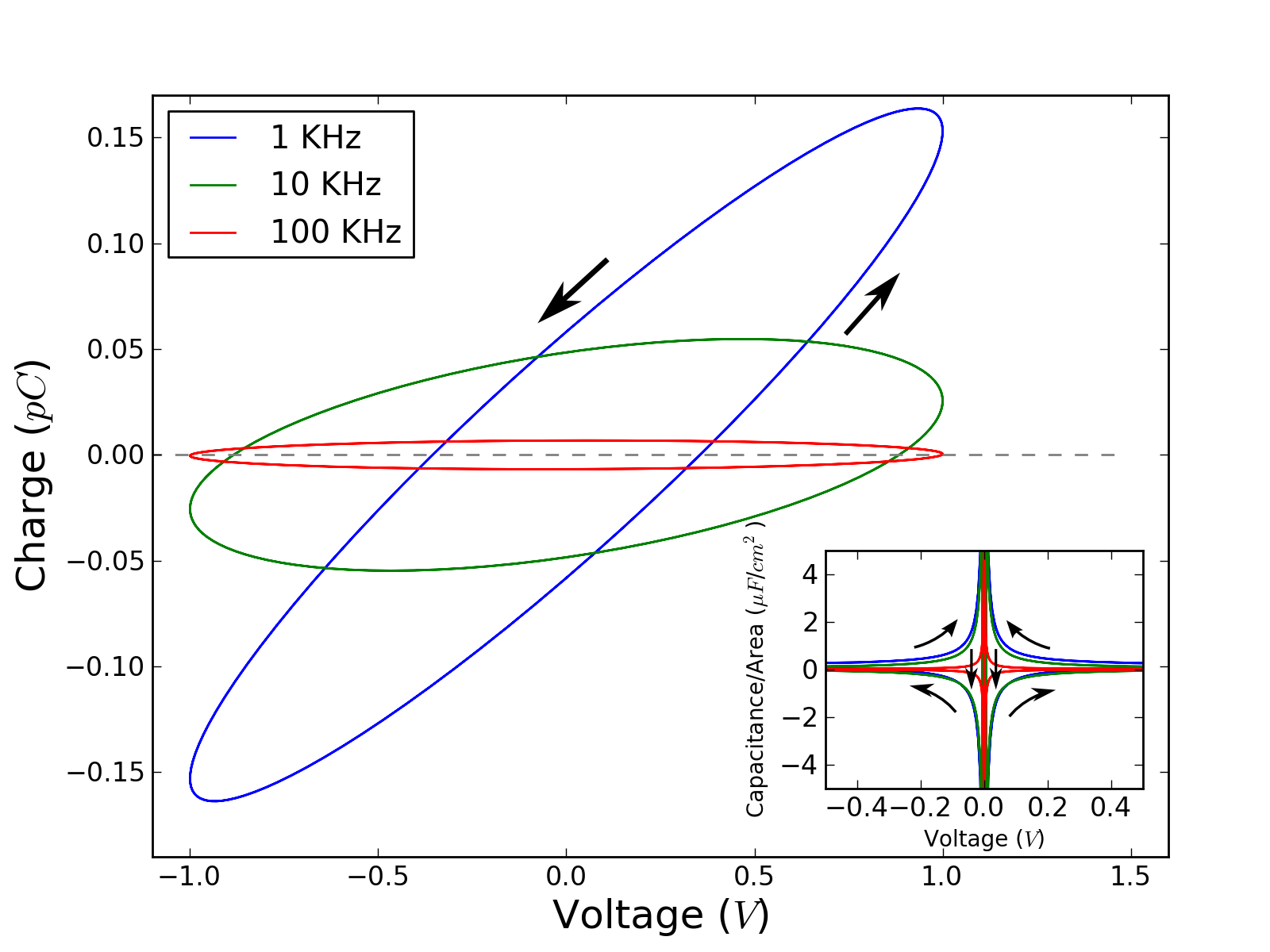}
\caption{(Color Online) Memcapacitive effects due to ionic transport across the pore. Main panel shows the charge on the external capacitor plates plotted versus the voltage across it for various frequencies of the electric field with $V_0=1.0~V$, $A=10~\mu m^2$, $L_p=25$~\AA, $r_p=7$~\AA, $\epsilon=7.5\epsilon_0$, $C_{ss}=\epsilon A/L_p$, and $C_{e-s}=10C_{ss}$ for a 1 M concentration of ions. The inset shows the capacitance plotted versus the voltage for the same frequencies. Memory
effects due to ionic transport across the pore occur at much lower frequencies than those due to the polarization of the ionic solution.
\label{fig:smallfreq}}
\end{figure}

{\it Memory at small frequencies} --- At small enough frequencies, there will be enough ion transport through the pore to play a role in the dynamics. In fact, it is no longer correct to assume that the charge associated with $C_{e-s}$ is the same as the charge associated with $C_{ss}$. Using Kirchoff's loop rule, we get an equation similar to Eq.~(\ref{voltage_drop}):
\begin{equation}
 2\frac{Q_{e-s}}{C_{e-s}}+2R_{s}\frac{dQ_{e-s}}{dt}+\frac{Q_{ss}}{C_{ss}}=V(t).\label{loop}
\end{equation}
except now we differentiate between the two charges, $Q_{e-s}$ (on the external plates) and $Q_{ss}$ (on the pore). Applying Kirchoff's junction rule we then get
\begin{equation}
 \frac{dQ_{e-s}}{dt}=\frac{dQ_{ss}}{dt}+\frac{V_{ss}}{R_{ss}}.\label{junction}
\end{equation}
where due to $C_{ss}$ and $R_{ss}$ being in parallel, $V_{ss}$ is simply the voltage on the $C_{ss}$ capacitor, $Q_{ss}/C_{ss}$, and we can approximate the conductance across the pore as $1/R_{ss}=2A_pen\mu/L_p$ where $A_p=\pi r_p^2$, with $r_p$ the pore radius, $e$ the electronic charge, $n$ the ion density, $\mu$ the ion mobility, and $L_p$ the length of the nanopore. We approximate $C_{ss}$ as that of a parallel plate capacitor, i.e., $C_{ss}=\epsilon A/L_p$ with $\epsilon$ the permittivity of Si$_3$N$_4$, and $A$ is again the area of the membrane surface. We can solve Eq.~\ref{loop} for $Q_{ss}$ and plug it into Eq.~\ref{junction} to arrive at the following second order equation for $Q_{e-s}$

\begin{eqnarray}
 2R_sC_{ss}\frac{d^2Q_{e-s}}{dt^2}  + \left(1+2\frac{C_{ss}}{C_{e-s}}+2\frac{R_s}{R_{ss}}\right)\frac{dQ_{e-s}}{dt} \nonumber \\ + 2\frac{Q_{e-s}}{R_{ss}C_{e-s}}
 =\frac{V(t)}{R_{ss}} + C_{ss}\frac{dV(t)}{dt}
\end{eqnarray}
For an ac field, this equation can be solved analytically but the solution is a bit involved. We then just plot it for different frequencies in Fig.~\ref{fig:smallfreq}. As anticipated,
we see that direct transport across the pore leads to an additional memory mechanism at specific frequencies, much smaller than the ones due to the polarization of the ionic solution.

{\it Experimental test} --- The results presented in this paper
show that there are two very distinct memory regimes for nanopores in solution. To observe the first memory mechanism due to the polarization of the ionic solution, ideally, one would need high frequencies (on the order
of 1GHz). On the other hand, much smaller frequencies (on the order of 1KHz) are necessary
to observe the additional memory effect due to transport through the pore. This indicates that the second memory mechanism would be the easiest one to detect.
However, in both cases we have shown that the capacitance acquires diverging
(and also negative) values. This result appears to hold also at frequencies that are relatively smaller than those required
to observe the first memory regime, namely at frequencies that are on the order of tens of MHz or less (see
Fig.~\ref{fig:infiniteC}). Therefore, with appropriate
control of the external circuit, at these frequencies one should at least be able to
observe non-trivial changes (manifested, e.g., in fast jumps
between positive and negative values) of the capacitance when the
bias crosses its zero value. We thus expect that by
simultaneously measuring the electric charge on the external electrodes
as shown in Fig.~\ref{fig:schematic}
(or, equivalently, the current at the electrodes) in the presence
of an applied time-dependent voltage should
allow a direct verification of our predictions for both types of memory.

{\it Conclusions} --- We have shown, using molecular dynamics
simulations, that a nanopore sequencing setup acts as a memcapacitor,
namely a capacitor with memory~\cite{diventra09}. The latter is due to two types of effects, and thus arises at very different frequencies of the
external bias. At high frequencies the finite
mobility of ions in water and hence the slow polarizability of the
ionic solution give rise to one type of memory. Memcapacitive effects, however, may also occur as a result of the ion transport through the pore at very low frequencies. These processes occur internally in the system
and, from the point of view of an external circuit, the whole system
behaves as an unusual capacitor. These effects may potentially
play a role in nanopore DNA sequencing proposals, especially those
based on ac-electric fields~\cite{sigalov08}, as well as in other
nanopore sensing applications. Moreover, the effect of the charge
buildup on the nanopore surface may influence DNA translocation
and its structure in proximity to the pore. Finally, due to the
ubiquitous nature of nanopores in biological processes, these
results may be relevant to specific ion dynamics when
time-dependent fields are of importance, such as in the action
potential formation and propagation during neuronal activity. We
thus hope this work will motivate studies in this direction.

We thank
Heiko Appel for useful discussions.
Financial support from the NIH-National Human Genome Research
Institute is gratefully acknowledged.

\bibliography{memcapacitor}
\end{document}